**Title: Nervus: A Comprehensive Deep Learning Classification, Regression, and Prognostication Tool for both Medical Image and Clinical Data Analysis**

Running title: Nervus: A Comprehensive Deep Learning Model for Radiology

Type: Technical Note/Special Reports

Toshimasa Matsumoto, MS, PhD[1,2], Shannon L Walston, MS[1], Yukio Miki, MD, PhD[1], Daiju Ueda, MD, PhD [1,2]

1) Department of Diagnostic and Interventional Radiology, Graduate School of Medicine, Osaka Metropolitan University, 1-4-3 Asahi-machi, Abeno-ku, Osaka 545-8585, Japan

2) Smart Life Science Lab, Center for Health Science Innovation, Osaka Metropolitan University, 1-4-3 Asahi-machi, Abeno-ku, Osaka 545-8585, Japan

**Corresponding author:** Daiju Ueda, MD, PhD

Department of Diagnostic and Interventional Radiology, Graduate School of Medicine, Osaka Metropolitan University, 1-4-3 Asahi-machi, Abeno-ku, Osaka 545-8585, Japan

Smart Life Science Lab, Center for Health Science Innovation

Phone: +81-6-6645-3831; Fax: +81-6-6646-6655

E-mail: ai.labo.ocu@gmail.com

**Abstract**

The goal of our research is to create a comprehensive and flexible library that is easy to use for medical imaging research, and capable of handling grayscale images, multiple inputs (both images and tabular data), and multi-label tasks. We have named it Nervus. Based on the PyTorch library, which is suitable for AI for research purposes, we created a four-part model to handle comprehensive inputs and outputs. Nervus consists of four parts. First is the dataloader, then the feature extractor, the feature mixer, and finally the classifier. The dataloader preprocesses the input data, the feature extractor extracts the features between the training data and ground truth labels, feature mixer mixes the features of the extractors, and the classifier classifies the input data from feature mixer based on the task. We have created Nervus, which is a comprehensive and flexible model library that is easy to use for medical imaging research which can handle grayscale images, multi-inputs and multi-label tasks. This will be helpful for researchers in the field of radiology.

**Keywords:**

Artificial Intelligence; Deep Learning; Classification; Regression; Prognostication; Radiology

**Abbreviations:**

AI: Artificial intelligence

CNN: Convolutional neural network

MLP: Multilayer perceptron

**Introduction**

In recent years, the performance of deep learning in computer vision tasks has revolutionized the field of imaging (1, 2). Radiology, a specialty based on images, has become the focus of many studies because of its affinity for deep learning (3, 4). This, coupled with the increasing demand for clinical imaging and the international shortage of radiologists (5, 6), has led to a surge of interest. As a result, there has been a significant increase in the number of deep learning applications in the field of radiology.

Among deep learning models, the most fundamental technique is the classification task (7-12). Other tasks include detection (13, 14) and segmentation (15, 16), the former of which can be regarded as region-based classification and the latter as pixel-based classification. Medical image classification tasks differ from general image classification tasks in four major ways. The first is the number of input image channels; 1 channel (grayscale) is common in medical image classification tasks, while 3 channels (RGB) are common for general images. Second, multi-input tasks are common in the medical field. For example, age, sex, lab data, etc. are often taken into account, not just images as in general image classification tasks (7). Third, multi-label tasks are common in the medical field. A single chest radiograph can contain multiple diseases at once, such as lung cancer and pneumonia. However, images in highly curated datasets such as ImageNet (17) usually have one label, so deep learning models should be modified to handle multi-label tasks. Fourth, disease features in medical images are often found in only a small portion of the image; ImageNet shows a large classification target in the middle of the picture, while a lung cancer on a chest radiograph is only a few centimeters in size at most. In such cases, image resizing should be done after much trial and error.

Compared with enriched deep learning libraries which can be used for general image classification, there are still few libraries widely available for medical image classification tasks. The goal of our research is to create a comprehensive and flexible library that is easy to use for medical imaging research, and capable of handling grayscale images, multiple inputs (both images and tabular data), and multi-label tasks; we have named it Nervus.

**Model framework**

**Overview**

Nervus is a library that can easily handle classification, regression, and prognostication models from simple images to more complex models combining images and tabular data. Nervus consists of four parts (Figure 1). First is the dataloader, then the feature extractor, the feature mixer, and finally the classifier. The dataloader pre-processes the input data, the feature extractor extracts the features between the training data and ground truth labels, the feature mixer mixes the features from the extractor, and the classifier classifies the input data based on tasks. Nervus can automatically assemble the structure of the model by providing variables for the task, the number of labels, the class of each label, neural networks, optimizer, criterion, epochs, batch size, augmentation, and input channels. It is also flexible in whether to use a pre-trained model, when to save the weight (each time the validation loss drops or only when it is the lowest), and whether to use a cpu or gpu as the processor (Table 1). Nervus is implemented with PyTorch and torchvision (18). All codes and tutorials are available on GitHub (https://github.com/Medical-AI-Lab/Nervus).

**Dataloader**

The dataloader pre-processes the input data. Specifically, it can convert the input channels in the image, format each batch size, set the augmentation, and determine the data sampling method. Nervus is designed to handle both 1 channel and 3 channel images. In medical imaging, 1 channel grayscale images are predominant, while most deep learning models are designed with 3 channels to handle RGB images. In medical imaging research, it is often desirable to use pre-trained models to improve accuracy and training stability. It may be not appropriate to transform a model pre-trained to use RGB to one which uses grayscale. Nervus can transform the weight of the first layer depending on the number of input channels. The dataloader also acts as a normalizer for tabular data when they are used as input. The augmentations available in torchvision can also be used. This makes it easier for the model to obtain stable training and high performance. Class imbalance is addressed during training using up-sampling. If you want to devise sampling for other tasks, you only need to change the sampler implementation without directly modifying the main implementation of the entire framework.

**Feature extractor**

The feature extractor is the most important part of deep learning, and extracts features between the input data and ground truth. An MLP is used as the feature extraction framework when the input is tabular data. An MLP is a forward propagating neural network in which each fully connected layer is followed by an activation function and a dropout layer. For example, clinical data such as age, gender, medical history, blood test results, etc. and ground truth features are extracted using an MLP. A CNN is used when the input is images. A CNN is a neural network which is used to recognize patterns in images. For example, image data such as chest radiographs, head MRI, and ground truth features are extracted using the CNN. The feature extractors used in Nervus, such as MLP and CNN, rely on torchvison, and the models implemented in it are easy to use. In addition, there is a file that defines only the feature extractor network, to which other PyTorch based models can be added, making it easy to implement the latest models that are released every day around the world.

**Feature mixer**

The feature mixer is a unique part of Nervus that is responsible for mixing the features from the feature extractors. In the case of tasks that deal with both image data and tabular data, by mixing the output of the feature extractors for both image and tabular data, a model can be created that captures the features between the image, tabular data, and ground truth. For example, when a physician reads a chest radiograph in a clinical setting, he/she rarely makes a diagnosis based on the image alone, but takes into account factors such as age. For example, if a chest radiograph shows a mass shadow, this raises the suspicion of a tumor of some kind. The older the patient is, the more malignancy is suspected, just as the younger the patient is, the less malignancy is suspected. Nervus can learn the relationship between both the tumor on the chest radiograph and the patient's age by using the feature mixer. If the input data is only tabular data or images, the feature mixer is not used.

**Classifier**

The classifier performs the role of classifying the outputs of the feature mixer (tabular and image inputs) or feature extractor (tabular or image input). This unit consists of one Fully Connected (FC) layer. FC layers in a neural network are those layers where all the inputs from one layer are connected to every activation unit of the next layer. Nervus is a library for multi-label tasks. In this section, we first review the terminology used for single-label and multi-label classification. In single-label classification, one label is assigned to a specific object. For example, this is the case with ImageNet and the MNIST tutorial. In the former case, one label out of 1000 classes is assigned to one general image. In the latter, one label is assigned to each handwritten character image from 10 classes ranging from 0 to 9. In multi-label classification, on the other hand, two or more labels can be assigned to a specific object at the same time. For example, the task is to determine whether a dog or a cat is in an image. In this case, it is possible for both a dog and a cat to appear in the image, one of them to appear, or neither of them to appear, which was not considered in the single-label case. In medicine, lung cancer and pneumonia can occur simultaneously in a single chest radiograph, so it is preferable to treat this as a multi-label task rather than a single-label task. At implementation of a multi-label task, the number of FC layers of the classifier is designed to be as many as the number of labels (≥1), arranged in parallel. The output from the feature mixer or feature extractor is the input to each FC layer. The total loss value is the sum of the loss values calculated for each of the labels in the classifier. In the case of a multi-label classification of lung cancer and pneumonia from chest radiographs, two FC layers are created because of the two labels, one FC layer is used to classify whether lung cancer is present or not, and the other FC layer is used to classify whether pneumonia is present or not. Of course, Nervus supports not only the multi-label classification task, but also the usual single-label classification task. For example, if we assume that a chest radiograph is used to classify the pathology of a lung cancer based on its shadow, a single-label task would be appropriate. In addition, the classifier can also perform a regression task by using MSE/RMSE/MAE as the loss function while it performs the classification task using cross entropy.

Nervus can also handle time-to-event prognostic models. Briefly, the Cox hazard model (19) is re-implemented using deep learning to overcome its original demerit of the assumption of linearity, and to be able to handle images and tabular data simultaneously. Specifically, a prognostic model can be constructed by providing images and/or tabular data as input, and by providing the binary classification of the event (e.g., death or recurrence) and the time to the event as ground truth. Using the results, Kaplan-Meier curves, log-rank tests, etc. can be performed. This is implemented based on the DeepSurv model (for tabular data) and previous research with CNN (for images) (20, 21). Specifically, we concatenated the output of the CNN to the fully connected layer of DeepSurv to create an end-to-end deep learning model.

**Discussion**

We have created a comprehensive and flexible model library that is easy to use for medical imaging research which can handle multi-input and multi-label tasks. This tool was created as an extension of a core technology in deep learning, the classification task. Further, it is closer to medical tasks in a clinical setting than most existing tools because physicians usually interpret images with other clinical data, and many patients have multiple diseases at once. Nervus is open source and publicly available and can be used freely for any purpose, although it was specifically created to be easy to use for radiology research.

To our knowledge, Nervus is the first comprehensive library to generate classification/regression/prognostication task models which can utilize medical images and/or tabular data. The interdisciplinary nature of the field makes it difficult to assemble a research team that brings together clinical, radiological, engineering, and computer science expertise (22), and so it is important to create a flexible tool that is easy for medical researchers to use, especially so for non-engineers. When using images for AI classification tasks (7-12), a large number of trials and errors are required to find the best hyperparameters. If the model is hard-coded, the implementation must be rewritten for each trial and error, which consumes a lot of time. When the model is created by adding tabular data as well as images as input, the number of trials will be even higher. In order to study this efficiently, Nervus has been soft-coded so that the various models and hyperparameters can be easily switched with python arguments. This makes it easy to use for beginners and saves a great deal of time and effort spent by researchers in medical AI research. It also facilitates not only classification tasks, but also regression and prognosis prediction. It can also be used as a base model for intermediate to higher level researchers as well, since the coding is refactored throughout, making it easy to use as a base model when adding new features.

The multi-label task implemented in Nervus is a task used when a single object can be classified into multiple classes simultaneously (23), and it also has two other advantages. First, it has advantages in terms of training time and model size. To create a model that can diagnose both lung cancer and pneumonia from a single chest radiograph using the conventional single-label model, we need to prepare radiographs which exclusively present lung cancer, pneumonia, or others (normal, etc.). Then, we have no choice but to create models for the two classification possibilities: "lung cancer vs. pneumonia + others" and "pneumonia vs. lung cancer + others". On the other hand, a multi-label task model can estimate the likelihood of lung cancer, pneumonia, and others all at once. Creating one model rather than two roughly halves the training time and the size of the model. Second, the multi-label model is advantageous for class imbalance problems when using biased training data (23). For example, if a model is created with many pneumonia images but insufficient lung cancer images, the model will successfully learn general pneumonia features during the training process because of the large amount of data. These features can be useful for lung cancer classification as well.

This tool has several limiting factors. First, detection (13, 14), segmentation (15, 16), super resolution (24), and image generation (25, 26) tasks are outside the scope of this tool. In addition, some of the code may be less readable or redundant due to the various features introduced to this model after the planning stage. Furthermore, the implementation is based on PyTorch (18) and cannot be handled by the TensorFlow library (27).

We have developed a modelling tool that can be flexibly applied to a variety of classification, regression, or prognostic tasks. These models are arranged based on the functions that we have added during the course of our daily research. We hope that these models will be of help for researchers in the field of radiology. This paper is an explanation of the technical aspects. Practical usage instructions can be found in the GitHub repository and Google Colaboratory.

**Figure legends**

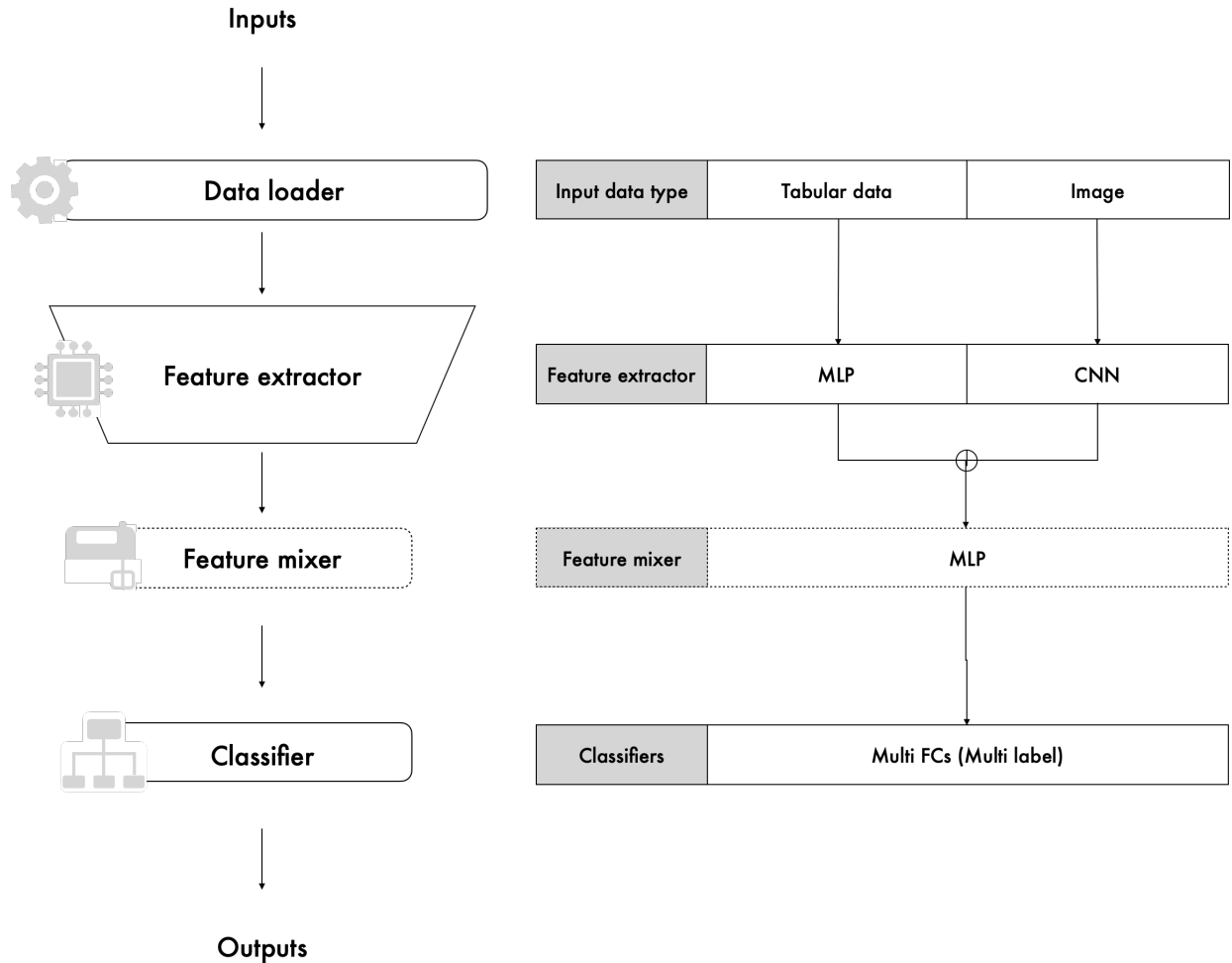

Figure 1: Model overview

Nervus consists of four parts. First is the dataloader, then the feature extractor, the feature mixer, and finally the classifier. The dataloader preprocesses the input data, the feature extractor extracts the features between the training data and ground truth labels, the feature mixer mixes the features of the extractors, and the classifier classifies the input data based on the task.

**Table**

Table 1: Model summary

|  | Classification task | Regression task | Prognostication task |
|---|---|---|---|
| Input data type | | | |
|    Image | ✓ | ✓ | ✓ |
|    Tabular data | ✓ | ✓ | ✓ |
| Output label type | | | |
|    Single label | ✓ | ✓ | ✓ |
|    Multi label | ✓ | ✓ | ✓ |
| Output class type | | | |
|    Binary | ✓ | - | ✓ |
|    Multi class | ✓ | - | - |
| Bit depth | | | |
|    8-bit | ✓ | ✓ | ✓ |
|    16-bit | ✓ | ✓ | ✓ |
| Color type | | | |
|    RGB | ✓ | ✓ | ✓ |
|    Grayscale | ✓ | ✓ | ✓ |
| Processor type | | | |
|    CPU | ✓ | ✓ | ✓ |
|    GPU | ✓ | ✓ | ✓ |
| Pretrained model | ✓ | ✓ | ✓ |